# Scalable van der Waals Photonics: High-Refractive-Index Gallium Sulfide Films with Single-Crystal Optical Properties

Aleksandr Slavich[1†], Georgy Ermolaev[1†], Dmitriy Grudinin[1†], Mikhail Tatmyshevskiy[2], Alexander Syuy[1], Nikolay Pak[1], Dmitry Yakubovsky[2], Mikhail Mironov[1], Adilet Toksumakov[1], Alexander Melentev[2], Elena Zhukova[2], Anton Minnekhanov[1], Gleb Tselikov[1], Ivan Kruglov[1], Sergey M. Novikov[2], Andrey Vyshnevyy[1], Aleksey Arsenin[1,2], Valentyn Volkov[1*]

[1]*Emerging Technologies Research Center, XPANCEO, Internet City, Emmay Tower, Dubai, United Arab Emirates*

[2]*Moscow Center for Advanced Studies, Kulakova str. 20, Moscow, 123592, Russia*

[†]These authors contributed equally to this work

*Correspondence should be addressed to the e-mail: vsv@xpanceo.com

## Abstract

The integration of high-refractive-index dielectrics into scalable photonic architectures is foundational to advancing integrated circuits and augmented reality (AR) displays. Van der Waals (vdW) materials offer exceptional optical properties, including high refractive indices and giant anisotropy, but their implementation is constrained by the small area and uncontrolled thickness of mechanically exfoliated flakes. Here, we demonstrate that atomic layer deposition (ALD) grown gallium sulfide (GaS) overcomes the trade-off between high optical performance and manufacturability, emerging as a large-scale vdW dielectric platform. Through rigorous optical and structural benchmarking against pristine single crystals, we establish that the optical constants ($n$, $k$) of ALD-GaS are virtually indistinguishable from single-crystal counterparts. By leveraging the retained out-of-plane anisotropy, we demonstrate that ALD-GaS enables superior suppression of crosstalk in densely integrated waveguides compared to conventional scalable high-index platforms. Our findings establish ALD-GaS as a technologically viable pathway for implementing anisotropic vdW materials in visible-spectrum photonics.

## Introduction

High-refractive-index (high-$n$) dielectrics have long served as the cornerstone of modern photonics, enabling the confinement and manipulation of light at the nanoscale[1–3]. The current material toolkit, while powerful, is a landscape of compromises[4,5]. Gallium nitride (GaN), a mature platform driven by the solid-state lighting industry, offers a high refractive index ($n \approx 2.3$) and a wide transparency window from the ultraviolet to the infrared, making it a cornerstone for light-emitting diodes[6] and integrated photonics[7]. Similarly, gallium phosphide (GaP) provides an exceptionally high index ($n > 3.0$), positioning it as a premier material for optical components in the visible spectrum[8,9]. Other established platforms, such as amorphous titanium dioxide ($TiO_2$) and lithium niobate, have become indispensable for meta-optics and active electro-optic modulation, respectively[10,11]. These materials are the workhorses for building photonic integrated circuits (PICs)[12], waveguides[13,14], metasurfaces and metamaterials[15], AR/VR displays[16,17], which demand transparent, high-index materials to efficiently guide and project light according to variable system requirements.

The emergence of two-dimensional (2D) and layered van der Waals (vdW) materials has opened a new frontier, offering a diverse palette of materials with high refractive index and anisotropic optical responses[18–22]. Yet, a fundamental chasm exists between the remarkable performance of small, mechanically exfoliated flakes and the requirements of manufacturable technology[23–25]. Current large-area synthesis methods often struggle to produce thickness-controllable films with the requisite crystal quality[26–30], leading to performance degradation from grain boundaries and defects, thus creating a critical bottleneck that impedes the translation of vdW photonics from the laboratory to real-world applications.

In this work, we report on the optical properties of large-area thick films of the layered semiconductor gallium sulfide (GaS), produced commercially via atomic layer deposition (ALD) on large area substrates. Using a combination of broadband spectroscopic ellipsometry, transmittance, and reflectance spectroscopy, we establish that the complex refractive index ($n$, $k$) of the ALD-grown GaS closely matches that of benchmark single crystals (mismatch less than 3% over broad UV-NIR range). Our work shows that ALD-GaS is transparent ($k < 0.001$) for wavelengths above ~480 nm, consistent with the material's wide bandgap of ~2.6 eV, and exhibits a high refractive index ($n$ reaches ≈2.8 within transparency window). This result effectively closes the gap between the ideal and the practical. The ALD-grown gallium sulfide film provides immediate access to the exceptional properties of vdW crystals for designing and prototyping the next generation of photonic components, from high-efficiency metalenses to compact waveguides, thereby paving the way for truly scalable van der Waals photonics.

## Results

**Structural and Morphological Characterization of ALD-grown GaS film**

To validate the quality of the commercially available thick gallium sulfide (GaS) film grown by Atomic Layer Deposition (ALD) (see Methods and Supplementary Information for details), we first investigate the structural properties. GaS crystallizes in a hexagonal structure (space group P6$_3$/mmc)[31], composed of S-Ga-Ga-S tetra-layers held together by weak vdW forces (**Fig. 1a**). As shown in **Fig. 1b**, a photograph of the ALD-grown GaS reveals a continuous, macroscopic foil-like film with millimetre-scale lateral dimensions. The film exhibits sufficient mechanical strength to undergo substrate release without fracture, which facilitates operation in suspended geometries or relocation onto other substrates.

We first verified the composition of the bulk film. Energy-dispersive X-ray (EDX) spectroscopy (**Fig. 1c**) confirmed high chemical purity and a near-ideal GaS stoichiometry, with an atomic ratio of Ga:S of 49.5:50.5. The minor Si peak originates from the underlying Si substrate. The inset scanning electron microscopy (SEM) image illustrates the film's morphology. As shown in **Fig. 1d,** optical profilometry measurements across the film edge confirmed the substantial thickness of approximately 20 μm (see inset in **Fig. 1d**).

Next, we assessed the crystalline quality of the ALD-grown GaS, a critical parameter often compromised by grain boundaries and defects during scalable synthesis. Raman spectroscopy (**Fig. 1e**), performed using 532 nm and 633 nm excitation wavelengths, revealed sharp, intense peaks corresponding to the characteristic vibrational modes of GaS, in good agreement with bulk GaS: $A^1_{1g}$ (188 cm$^{-1}$), $E^1_{2g}$ (295 cm$^{-1}$), and $A^2_{1g}$ (360 cm$^{-1}$)[32]. The narrow linewidths and the absence of additional peaks associated with secondary phases or amorphous by-products indicate high crystalline quality and phase purity of the ALD-grown films.

The layered vdW structure and high quality of the bulk film enable mechanical exfoliation of thin flakes onto a substrate. **Fig. 1f** demonstrates representative flakes obtained by mechanical exfoliation from a bulk ALD-grown film. Atomic force microscopy (AFM) analysis of an exfoliated flake surface (**Fig. 1g**) reveals an atomically flat topography, evidenced by an ultra-low root-mean-square (RMS) roughness of 0.14 nm over the scanned area.

Finally, we confirmed the crystal quality of GaS flakes by transmission electron microscopy (TEM). A high-resolution TEM image (**Fig. 1h**) displays clear lattice fringes with an interplanar spacing of 0.32 nm, consistent with d-spacings in GaS[31], and the corresponding fast Fourier transform (FFT) shows sharp diffraction spots (inset of **Fig. 1h**). Furthermore, the selected area electron diffraction (SAED) pattern (**Fig. 1i**) exhibits a distinct hexagonal pattern along the [001] direction. Collectively, these results demonstrate that ALD-grown GaS forms thick films with high crystal quality and stoichiometric composition over large areas.

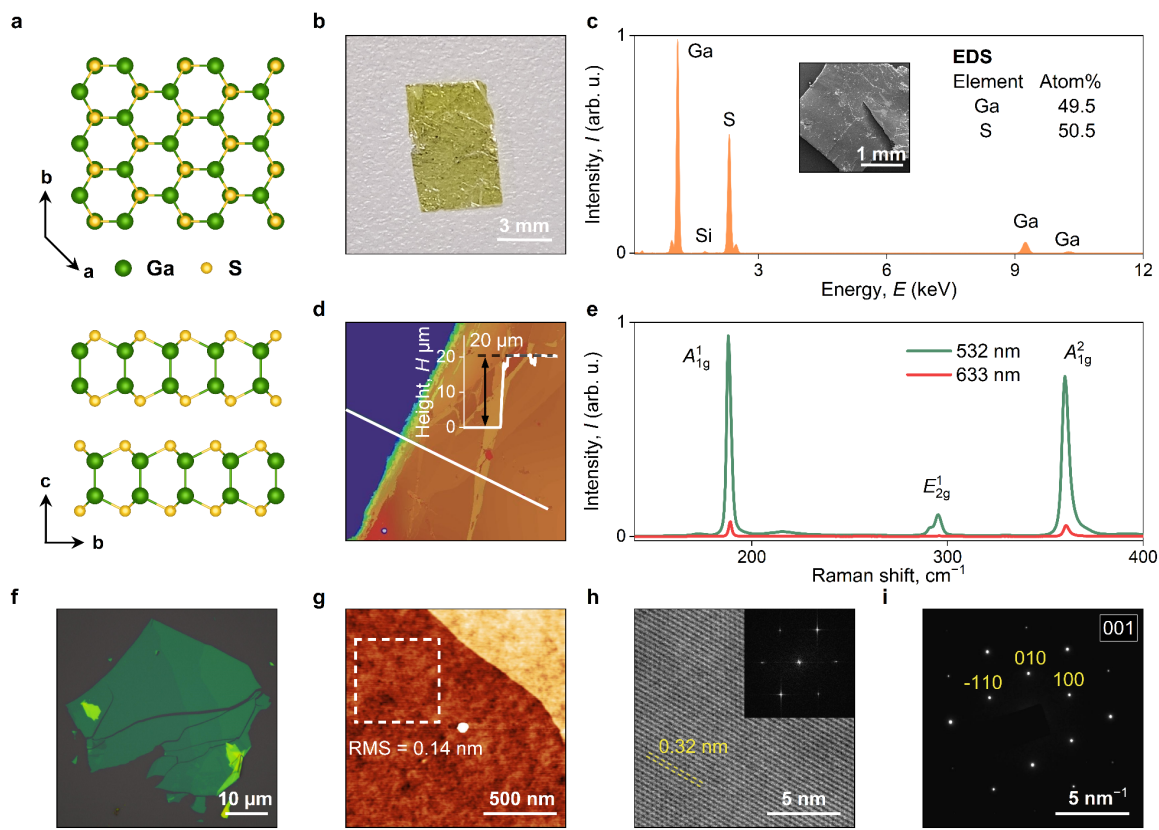

**FIG. 1.** Structural and morphological characterization of ALD-grown GaS. **(a)** Top-down and side views of the GaS crystal structure. **(b)** Optical image of commercial ALD-grown thick GaS film. **(c)** Energy-dispersive X-ray (EDX) spectrum of a thick ALD-grown GaS film transferred onto a Si substrate. The inset shows a scanning electron microscopy (SEM) image of the film. **(d)** Image of the film edge from an optical profilometer. The inset shows the height profile along the solid white line. **(e)** Raman spectra of the film acquired with 532 nm (green) and 633 nm (red) laser excitation. **(f)** Optical micrograph of a GaS flake exfoliated and transferred onto a substrate. **(g)** Atomic force microscopy (AFM) image of the flake surface. The dashed square indicates the area used for the root-mean-square (RMS) roughness calculation. **(h)** High-resolution transmission electron microscopy (HR-TEM) image of a flake transferred onto a TEM grid. Inset: Corresponding Fast Fourier Transform (FFT). **(i)** Selected-area electron diffraction (SAED) pattern from the same flake.

**Optical Properties of ALD-Grown GaS**

Having established the high crystalline quality of the ALD-grown GaS, we proceeded to characterize its optical properties. We first evaluated the material's transparency window by measuring the optical transmittance of a suspended 20-μm-thick film (**Fig. 2a**). The film is transparent in the near-infrared region, with the absorption edge located in the visible spectrum. A corresponding Tauc analysis (**Fig. 2a**, inset) confirms an indirect bandgap ($E_g$) of approximately 2.6 eV, aligning well with established values for bulk GaS[33].

To accurately determine the optical constants $n(\lambda)$ and $k(\lambda)$ across a broad spectral range, we performed spectroscopic ellipsometry characterization of the 20-μm-thick GaS film in combination with broadband transmittance ($T$) and reflectance ($R$) measurements (see Supplementary Note 1,2). **Fig. 2b** shows representative experimental transmittance ($T$) and reflectance ($R$) spectra for a flake exfoliated from the film (solid lines). The dashed lines show spectra calculated using a best-fit optical model derived from the material's optical constants, demonstrating excellent agreement between the experimental data and the model.

The resulting optical constants of the ALD-grown GaS are presented in **Fig. 2c**. The material exhibits a high refractive index and measurable out-of-plane birefringence, characteristic of the layered GaS structure[18]. The in-plane refractive index $n_{ab}$ is notably high, exceeding ~2.8 throughout the visible and near-infrared spectrum. Concurrently, the in-plane extinction coefficient $k_{ab}$ rapidly diminishes above the band edge, confirming negligible absorption within the transparency window. These experimental findings are in good agreement with our *ab initio* density functional theory (DFT) calculations (**Fig. 2c**, dotted lines).

Crucially, we benchmarked the optical performance of the ALD-grown GaS against established values for high-quality mechanically exfoliated single crystals[33]. The comparison reveals that the refractive index (**Fig. 2d**) and the extinction coefficient (**Fig. 2e**) of the ALD-synthesized GaS are virtually indistinguishable from those of their single-crystal counterparts. The refractive index mismatch is less than 3% in the NIR range, decreasing towards the absorption edge. This comparison indicates that the ALD process can yield large-area vdW films that replicate the benchmark optical performance of exfoliated single crystals, effectively closing the gap between idealized material properties and scalable photonic integration.

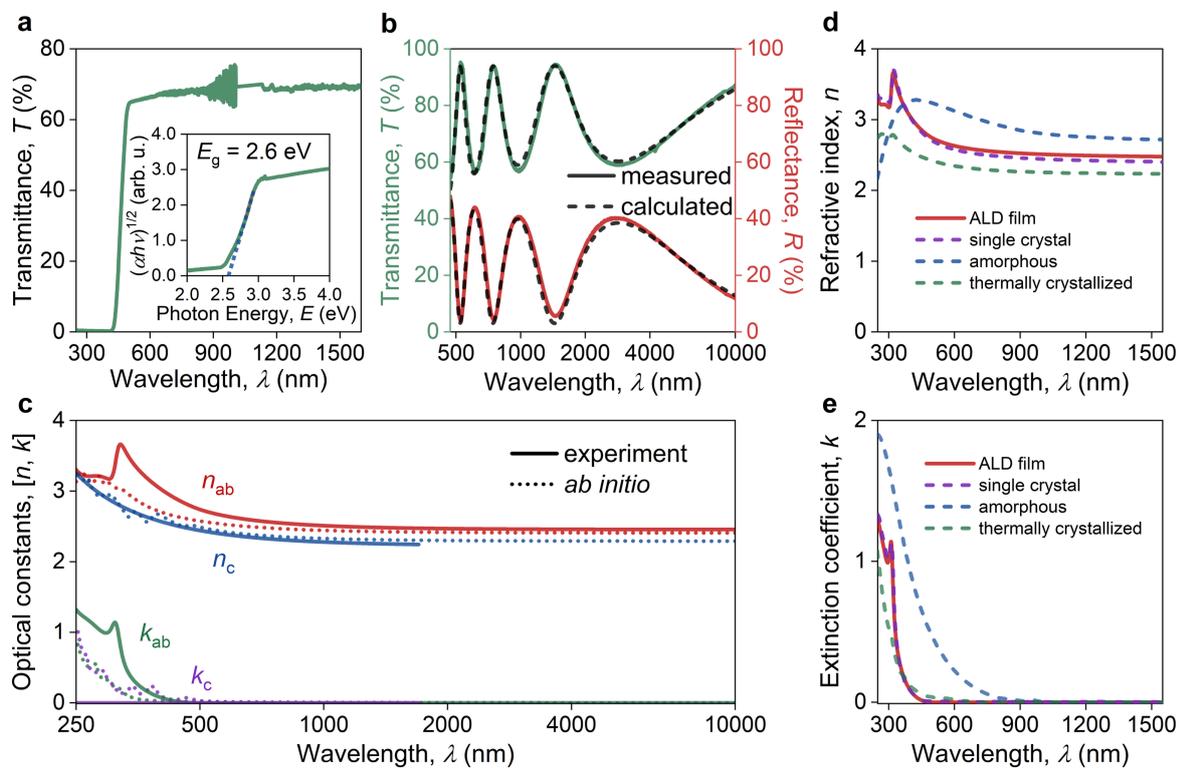

**FIG. 2.** Optical properties of ALD-grown GaS. **(a)** Transmittance spectrum of suspended 20μm-thick GaS film and its corresponding Tauc plot (inset). **(b)** Representative transmittance (*T*) and reflectance (*R*) spectra of a flake obtained by exfoliation from a thick GaS film. The dashed lines show the results of modeling based on the optical constants. **(c)** The resulting optical constants of ALD-grown GaS. The dotted lines represent calculations from first principles. **(d)**, **(e)** Comparison of the obtained in-plane **(d)** refractive index and **(e)** extinction coefficient with the optical constants of single-crystal flakes[33], amorphous and thermally crystallized films[34].

## Prospects for High-Density Photonic Integration

The realization of large-area ALD-grown GaS films that retain the benchmark optical properties of single crystals offers a promising route toward scalable, high-density photonic integrated circuits. A primary challenge in dense integration is minimizing evanescent coupling, or crosstalk, between adjacent components. The high refractive index and inherent optical anisotropy of vdW materials provides a distinct advantage over conventional isotropic platforms by enabling enhanced light confinement and suppressed inter-waveguide coupling[35–38].

We simulated the propagation distance required for complete power transfer between coupled waveguides (crosstalk length, $L_{ct}$) for several well-known vdW materials, focusing on the impact of the out-of-plane refractive index $n_c$. For each value of $n_c$ we have optimized the waveguide dimensions to achieve maximal crosstalk distance within the single-mode regime. **Fig. 3a** illustrates the maximum $L_{ct}$ (at $\lambda_{k<0.001}$) for GaS and other vdW semiconductors. The simulations confirm that increasing anisotropy suppresses crosstalk: $L_{ct}$ increases substantially as $n_c$ decreases relative to the in-plane index $n_{ab}$. For GaS, its intrinsic anisotropy yields an $L_{ct}$ significantly longer than that of a hypothetical isotropic equivalent (with $n_c = n_{ab}$), demonstrating the critical benefit of its preserved layered structure.

Having established the advantage conferred by the anisotropy preserved in the scalable ALD-grown GaS, we benchmarked its performance against established, commercially scalable photonic platforms, such as Silicon (Si)[39] and Titanium Dioxide (TiO$_2$). **Fig. 3b** presents the calculated cross-talk length $L_{ct}$ as a function of the inter-waveguide separation distance ($d$), with the waveguide geometries optimized for each configuration at $\lambda_{k<0.001}$.

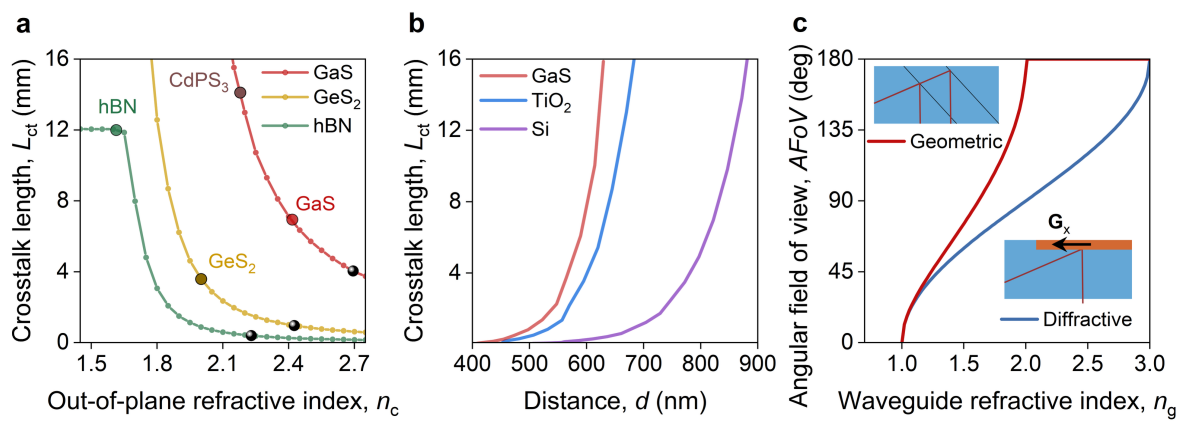

**FIG. 3.** Prospective applications of ALD-grown GaS. **(a)** Maximum cross-talk length in the single-mode regime as a function of the out-of-plane refractive index for different representative materials: hBN[37], GeS$_2$[40], CdPS$_3$[41]. Black dots indicate $n_{\text{out-of-plane}} = n_{\text{in-plane}}$. The circles mark points calculated at actual $n_c$ of the corresponding materials. **(b)** Relation of crosstalk length $L_{ct}$ and distance $d$ between cores of the waveguides for high-refractive index scalable materials. The width is optimized for each point. **(c)** Angular field of view of AR/VR waveguide displays with geometric (transparent mirror-based) and diffractive (diffractive/holographic optical element-based) outcouplers as a function of the refractive index of the waveguide $n_g$, calculated using the analytical expressions described in Supplementary Note 3.

The results demonstrate a clear performance advantage for the GaS platform. Owing to its unique combination of a high in-plane refractive index (facilitating tight mode confinement) and significant anisotropy (suppressing crosstalk), preserved across large areas during synthesis process, GaS consistently outperforms Si and TiO$_2$. For any given separation distance, GaS exhibits a substantially longer $L_{ct}$. Conversely, to achieve a target level of optical isolation (e.g., $L_{ct}$ = 10 mm), GaS waveguides can be integrated with a smaller pitch.

Another promising application venue for GaS is AR/VR optics. Waveguide-based augmented reality displays require high-refractive-index materials to achieve a wide angular field of view (AFoV)[16]. Simple estimation for the waveguide with the reflective light couplers also called "geometric waveguides"[42] give $AFoV = 2arcsin\sqrt{(n_g^2 - n_g)/2}$ for $n_g$ < 2 where $n_g$ is the waveguide refractive index and $AFoV = 2arcsin\sqrt{(n_g - 1)/2}$ for $n_g$ < 3 for waveguides with holographic and diffractive couplers (see **Fig. 3c** and Supplementary Note 3). While GaS is not suitable for full-color devices due to the strong optical absorption of blue light, its scalability makes it an ideal material for prototyping highly refractive optics, and AR waveguides in particular, and monochrome devices.

## Conclusions

We established atomic layer deposition (ALD) grown gallium sulfide (GaS) as a technologically viable, high-refractive-index dielectric platform that unites the attractive optical properties of van der Waals (vdW) materials with the imperatives of scalable manufacturing. Our findings demonstrate that the optical constants of ALD-GaS films closely match those of pristine single crystals, confirming the preservation of optical properties at large scale. The material anisotropy which is conserved during ALD growth provides a fundamental advantage over conventional isotropic materials, enabling enhanced optical confinement and suppressed crosstalk, thereby facilitating denser photonic integration. The realization of benchmark-quality vdW optics in a scalable format resolves the long-standing trade-off between performance and manufacturability, providing a foundation for the development of next-generation visible-spectrum photonic integrated circuits, ultra-compact optical components, and advanced display technologies.

## Supplementary Information

Supplementary Information contains sections Materials and Methods, Additional Figures, and tabulated optical constants.

## Competing Interests

The authors declare no competing interests.

## Acknowledgments

The authors thank Dr. Davit Ghazaryan for fruitful discussion. M.T., S.M.N., and A.A. acknowledge the financial support from the RSF (Grant No. 25-19-00326).

## Data Availability

The data that support the findings of this study are available from the corresponding author upon reasonable request.

**References**

[1] J.B. Khurgin, "How to deal with the loss in plasmonics and metamaterials," Nat. Nanotechnol. **10**(1), 2–6 (2015).

[2] Y. Kivshar, "All-dielectric meta-optics and non-linear nanophotonics," Natl. Sci. Rev. **5**(2), 144–158 (2018).

[3] V.E. Babicheva, and A.B. Evlyukhin, "Mie-resonant metaphotonics," Adv. Opt. Photonics **16**(3), 539 (2024).

[4] J.B. Khurgin, "Expanding the Photonic Palette: Exploring High Index Materials," ACS Photonics **9**(3), 743–751 (2022).

[5] D.G. Baranov, D.A. Zuev, S.I. Lepeshov, O.V. Kotov, A.E. Krasnok, A.B. Evlyukhin, and B.N. Chichkov, "All-dielectric nanophotonics: the quest for better materials and fabrication techniques," Optica, OPTICA **4**(7), 814–825 (2017).

[6] H. Wu, X. Lin, Q. Shuai, Y. Zhu, Y. Fu, X. Liao, Y. Wang, Y. Wang, C. Cheng, Y. Liu, L. Sun, X. Luo, X. Zhu, L. Wang, Z. Li, X. Wang, D. Li, and A. Pan, "Ultra-high brightness Micro-LEDs with wafer-scale uniform GaN-on-silicon epilayers," Light Sci. Appl. **13**(1), 284 (2024).

[7] C. Xiong, W. Pernice, K.K. Ryu, C. Schuck, K.Y. Fong, T. Palacios, and H.X. Tang, "Integrated GaN photonic circuits on silicon (100) for second harmonic generation," Opt. Express **19**(11), 10462–10470 (2011).

[8] D.J. Wilson, K. Schneider, S. Hönl, M. Anderson, Y. Baumgartner, L. Czornomaz, T.J. Kippenberg, and P. Seidler, "Integrated gallium phosphide nonlinear photonics," Nat. Photonics **14**(1), 57–62 (2020).

[9] V.V. Fedorov, O.Y. Koval, D.R. Ryabov, S.V. Fedina, I.E. Eliseev, D.A. Kirilenko, D.A. Pidgayko, A.A. Bogdanov, Y.M. Zadiranov, A.S. Goltaev, G.A. Ermolaev, A.V. Arsenin, S.V. Makarov, A.K. Samusev, V.S. Volkov, and I.S. Mukhin, "Nanoscale gallium phosphide epilayers on sapphire for low-loss visible nanophotonics," ACS Appl. Nano Mater. **5**(7), 8846–8858 (2022).

[10] S. Sun, Z. Zhou, C. Zhang, Y. Gao, Z. Duan, S. Xiao, and Q. Song, "All-dielectric full-color printing with $TiO_2$ metasurfaces," ACS Nano **11**(5), 4445–4452 (2017).

[11] Z. Li, R.N. Wang, G. Lihachev, J. Zhang, Z. Tan, M. Churaev, N. Kuznetsov, A. Siddharth, M.J. Bereyhi, J. Riemensberger, and T.J. Kippenberg, "High density lithium niobate photonic integrated circuits," Nat. Commun. **14**(1), 4856 (2023).

[12] P.I. Borel, B. Bilenberg, L.H. Frandsen, T. Nielsen, J. Fage-Pedersen, A.V. Lavrinenko, J.S. Jensen, O. Sigmund, and A. Kristensen, "Imprinted silicon-based nanophotonics," Opt. Express **15**(3), 1261–1266 (2007).

[13] L.H. Frandsen, A.V. Lavrinenko, J. Fage-Pedersen, and P.I. Borel, "Photonic crystal waveguides with semi-slow light and tailored dispersion properties," Opt. Express **14**(20), 9444–9450 (2006).

[14] R. Jacobsen, A. Lavrinenko, L. Frandsen, C. Peucheret, B. Zsigri, G. Moulin, J. Fage-Pedersen, and P. Borel, "Direct experimental and numerical determination of extremely high group indices in photonic crystal waveguides," Opt. Express **13**(20), 7861–7871 (2005).

[15] M. Khorasaninejad, A.Y. Zhu, C. Roques-Carmes, W.T. Chen, J. Oh, I. Mishra, R.C. Devlin, and F. Capasso, "Polarization-insensitive metalenses at visible wavelengths," Nano Lett. **16**(11), 7229–7234 (2016).

[16] Z. Li, P. Lin, Y.-W. Huang, J.-S. Park, W.T. Chen, Z. Shi, C.-W. Qiu, J.-X. Cheng, and F. Capasso, "Meta-optics achieves RGB-achromatic focusing for virtual reality," Sci. Adv. **7**(5), eabe4458 (2021).

[17] J.P. Rolland, and J. Goodsell, "Waveguide-based augmented reality displays: a highlight," Light Sci. Appl. **13**(1), 22 (2024).

[18] G.A. Ermolaev, D.V. Grudinin, Y.V. Stebunov, K.V. Voronin, V.G. Kravets, J. Duan, A.B. Mazitov, G.I. Tselikov, A. Bylinkin, D.I. Yakubovsky, S.M. Novikov, D.G. Baranov, A.Y. Nikitin, I.A. Kruglov, T. Shegai, P. Alonso-González, A.N. Grigorenko, A.V. Arsenin, K.S. Novoselov, and V.S. Volkov, "Giant optical anisotropy in transition metal dichalcogenides for next-generation photonics," Nat. Commun. **12**(1), 854 (2021).

[19] B. Munkhbat, B. Küçüköz, D.G. Baranov, T.J. Antosiewicz, and T.O. Shegai, "Nanostructured transition metal dichalcogenide multilayers for advanced nanophotonics," Laser Photon. Rev. **17**(1),

2200057 (2023).

[20] P.G. Zotev, P. Bouteyre, Y. Wang, S.A. Randerson, X. Hu, L. Sortino, Y. Wang, T. Shegai, S.-H. Gong, A. Tittl, I. Aharonovich, and A.I. Tartakovskii, "Nanophotonics with multilayer van der Waals materials," Nat. Photonics **19**(8), 788–802 (2025).

[21] B. Munkhbat, P. Wróbel, T.J. Antosiewicz, and T.O. Shegai, "Optical constants of several multilayer transition metal dichalcogenides measured by spectroscopic ellipsometry in the 300-1700 nm range: High index, anisotropy, and hyperbolicity," ACS Photonics **9**(7), 2398–2407 (2022).

[22] T.D. Green, D.G. Baranov, B. Munkhbat, R. Verre, T. Shegai, and M. Käll, "Optical material anisotropy in high-index transition metal dichalcogenide Mie nanoresonators," Optica **7**(6), 680 (2020).

[23] C. Gautam, B. Thakurta, M. Pal, A.K. Ghosh, and A. Giri, "Wafer scale growth of single crystal two-dimensional van der Waals materials," Nanoscale **16**(12), 5941–5959 (2024).

[24] T.-A. Chen, C.-P. Chuu, C.-C. Tseng, C.-K. Wen, H.-S.P. Wong, S. Pan, R. Li, T.-A. Chao, W.-C. Chueh, Y. Zhang, Q. Fu, B.I. Yakobson, W.-H. Chang, and L.-J. Li, "Wafer-scale single-crystal hexagonal boron nitride monolayers on Cu (111)," Nature **579**(7798), 219–223 (2020).

[25] G. Xue, B. Qin, C. Ma, P. Yin, C. Liu, and K. Liu, "Large-area epitaxial growth of transition metal dichalcogenides," Chem. Rev. **124**(17), 9785–9865 (2024).

[26] X. Zhao, Y. Ji, J. Chen, W. Fu, J. Dan, Y. Liu, S.J. Pennycook, W. Zhou, and K.P. Loh, "Healing of planar defects in 2D materials via grain boundary sliding," Adv. Mater. **31**(16), e1900237 (2019).

[27] M.G. Kozodaev, A.S. Slavich, R.I. Romanov, S.S. Zarubin, and A.M. Markeev, "Influence of reducing agent on properties of thin $WS_2$ nanosheets prepared by sulfurization of atomic layer-deposited $WO_3$," J. Phys. Chem. C Nanomater. Interfaces **124**(51), 28169–28177 (2020).

[28] G.A. Ermolaev, D.I. Yakubovsky, M.A. El-Sayed, M.K. Tatmyshevskiy, A.B. Mazitov, A.A. Popkova, I.M. Antropov, V.O. Bessonov, A.S. Slavich, G.I. Tselikov, I.A. Kruglov, S.M. Novikov, A.A. Vyshnevyy, A.A. Fedyanin, A.V. Arsenin, and V.S. Volkov, "Broadband optical constants and nonlinear properties of SnS2 and SnSe2," Nanomaterials (Basel) **12**(1), 141 (2021).

[29] G.A. Ermolaev, K.V. Voronin, M.K. Tatmyshevskiy, A.B. Mazitov, A.S. Slavich, D.I. Yakubovsky, A.P. Tselin, M.S. Mironov, R.I. Romanov, A.M. Markeev, I.A. Kruglov, S.M. Novikov, A.A. Vyshnevyy, A.V. Arsenin, and V.S. Volkov, "Broadband optical properties of atomically thin PtS2 and PtSe2," Nanomaterials (Basel) **11**(12), 3269 (2021).

[30] C. Liu, T. Liu, Z. Zhang, Z. Sun, G. Zhang, E. Wang, and K. Liu, "Understanding epitaxial growth of two-dimensional materials and their homostructures," Nat. Nanotechnol. **19**(7), 907–918 (2024).

[31] A. Kuhn, A. Chevy, and R. Chevalier, "Refinement of the 2*H* GaS β-type," Acta Crystallogr. B **32**(3), 983–984 (1976).

[32] D.J. Late, B. Liu, H.S.S.R. Matte, C.N.R. Rao, and V.P. Dravid, "Rapid characterization of ultrathin layers of chalcogenides on $SiO_2$/Si substrates," Adv. Funct. Mater. **22**(9), 1894–1905 (2012).

[33] Y. Gutiérrez, D. Juan, S. Dicorato, G. Santos, M. Duwe, P.H. Thiesen, M.M. Giangregorio, F. Palumbo, K. Hingerl, C. Cobet, P. García-Fernández, J. Junquera, F. Moreno, and M. Losurdo, "Layered gallium sulfide optical properties from monolayer to CVD crystalline thin films," Opt. Express **30**(15), 27609–27622 (2022).

[34] Y. Gutiérrez, S. Dicorato, A.P. Ovvyan, F. Brückerhoff-Plückelmann, J. Resl, M.M. Giangregorio, K. Hingerl, C. Cobet, M. Schiek, M. Duwe, P.H. Thiesen, W.H.P. Pernice, and M. Losurdo, "Layered Gallium Monosulfide as Phase-Change Material for Reconfigurable Nanophotonic Components On-Chip," Advanced Optical Materials **12**(3), (2024).

[35] G. Ermolaev, D. Grudinin, K. Voronin, A. Vyshnevyy, A. Arsenin, and V. Volkov, "Van Der Waals Materials for Subdiffractional Light Guidance," Photonics **9**(10), 744 (2022).

[36] A.A. Vyshnevyy, G.A. Ermolaev, D.V. Grudinin, K.V. Voronin, I. Kharichkin, A. Mazitov, I.A. Kruglov, D.I. Yakubovsky, P. Mishra, R.V. Kirtaev, A.V. Arsenin, K.S. Novoselov, L. Martin-Moreno, and V.S. Volkov, "van der Waals Materials for Overcoming Fundamental Limitations in Photonic Integrated Circuitry," Nano Lett. **23**(17), 8057–8064 (2023).

[37] D.V. Grudinin, G.A. Ermolaev, D.G. Baranov, A.N. Toksumakov, K.V. Voronin, A.S. Slavich, A.A. Vyshnevyy, A.B. Mazitov, I.A. Kruglov, D.A. Ghazaryan, A.V. Arsenin, K.S. Novoselov, and V.S. Volkov,

"Hexagonal boron nitride nanophotonics: a record-breaking material for the ultraviolet and visible spectral ranges," Mater Horiz **10**(7), 2427–2435 (2023).

[38] H. Ling, R. Li, and A.R. Davoyan, "All van der Waals integrated nanophotonics with bulk transition metal dichalcogenides," ACS Photonics **8**(3), 721–730 (2021).

[39] C.M. Herzinger, B. Johs, W.A. McGahan, J.A. Woollam, and W. Paulson, "Ellipsometric determination of optical constants for silicon and thermally grown silicon dioxide via a multi-sample, multi-wavelength, multi-angle investigation," Journal of Applied Physics **83**(6), 3323–3336 (1998).

[40] A.S. Slavich, G.A. Ermolaev, I.A. Zavidovskiy, D.V. Grudinin, K.V. Kravtsov, M.K. Tatmyshevskiy, M.S. Mironov, A.N. Toksumakov, G.I. Tselikov, I.M. Fradkin, K.V. Voronin, M.R. Povolotskiy, O.G. Matveeva, A.V. Syuy, D.I. Yakubovsky, D.M. Tsymbarenko, I. Kruglov, D.A. Ghazaryan, S.M. Novikov, A.A. Vyshnevyy, A.V. Arsenin, V.S. Volkov, and K.S. Novoselov, "Germanium disulfide as an alternative high refractive index and transparent material for UV-visible nanophotonics," Light Sci. Appl. **14**(1), 213 (2025).

[41] M.R. Povolotskiy, A.S. Slavich, G.A. Ermolaev, D.V. Grudinin, N.V. Pak, I.A. Zavidovskiy, K.V. Kravtsov, A.A. Minnekhanov, M.K. Tatmyshevskiy, A.V. Syuy, D.I. Yakubovsky, A. Mazitov, L.A. Klimova, A.N. Toksumakov, A.V. Melentev, E. Zhukova, D.A. Ghazaryan, G. Tselikov, I. Kruglov, S.M. Novikov, A.A. Vyshnevyy, A.V. Arsenin, K.S. Novoselov, and V.S. Volkov, "Record Index-Bandgap Trade-off: CdPS3 as a High-Index van der Waals Platform for Ultraviolet-Visible Nanophotonics," arXiv 2511.14269, (2025).

[42] Y. Ding, Q. Yang, Y. Li, Z. Yang, Z. Wang, H. Liang, and S.-T. Wu, "Waveguide-based augmented reality displays: perspectives and challenges," eLight **3**, article number 24, (2023).